\documentclass[12pt]{article}

\usepackage{amssymb}

\usepackage{amsmath}

\begin{document}

\baselineskip=18truept

\title{Sum rules and the domain after the last node of an eigenstate}
\author{C.V.Sukumar\\
Department of Physics,University of Oxford\\
Theoretical Physics, 1 Keble Road, Oxford OX1 3NP}
\maketitle

\begin{abstract}
It is shown that it is possible to establish sum rules that must be
satisfied at the nodes and extrema of the eigenstates of confining
potentials which are functions of a single variable. At any boundstate energy the Schroedinger equation has two linearly
independent solutions one of which is normalisable while the other
is not. In the domain
after the last node of a boundstate eigenfunction the unnormalisable
linearly independent solution has a simple form which enables the
construction of functions analogous to Green's functions that lead to certain 
sum rules. One set of sum rules give conditions
that must be satisfied at the nodes
and extrema of the boundstate eigenfunctions of confining
potentials. Another sum rule establishes a relation between an
integral involving an eigenfunction in the domain after the last node
and a sum involving all the 
eigenvalues and eigenstates. Such sum rules may be useful in the study of properties of confining
potentials. The exactly solvable cases of the particle in a box and the
simple harmonic oscillator are used to
illustrate the procedure. The relations between one of the sum rules and
two-particle densities and a construction based on Supersymmetric Quantum Mechanics are discussed.
\end{abstract}
\noindent

PACS: 02.30.Gp, 03.65.-w, 11.30.Pb

\vfill\eject
\noindent

\section{\noindent\textbf{Introduction}}
There is a well defined procedure for constructing Green's functions for
describing solutions to linear second order differential equations with inhomogeneous
terms (Morse and Feshbach 1953 [1]). This procedure may be employed to study solutions to the
Schroedinger equation in one dimension. For problems with spherical
symmetry partial wave decomposition effectively reduces the
three-dimensional Schroedinger equation to a radial equation in r-space
and hence the techniques for constucting Green's functions in
one-dimension are applicable. The Green's functions may be used to
establish trace formulae in which the integrals over the Green's
function may be related to sums involving the eigenvalues of the
homogeneous differential equation (Berry 1986 [2], Sukumar 1990 [3], Voros 2000 [4]). Such trace formulae
are very useful in checking the accuracy of the numerically computed
eigenvalues. The recent interest in the real spectra of non-hermitian
Hamiltonians exhibiting PT symmetry (Bender and Boettcher 1998 [5],
Mezincescu 2000 [6], Bender and Wong 2001 [7]) has initiated accurate numerical computation of
eigenvalues of PT symmetric Hamiltonians and the trace formulae have
proved to be useful. 

In this work we consider the two
linearly independent solutions to the Schroedinger equation at an eigenenergy
 and show that it is possible to constuct functions
 which are suitable for
studying various sums over eigenstates in the domain outside the last
node of a chosen eigenfunction. In section 2 it is shown that it is possible to
construct new sum rules involving all the eigenstates and
eigenvalues. In section 3 the
examples of a particle in a box and the simple harmonic oscillator are
used to illustrate the sum rules. The relations between the sum rules,
two particle densities and Super Symmetric Quantum Mechanics are
discussed in section 4.
Units in which $\hbar =1$ and the mass $\mu =\frac{1}{2}$ are used
throughout the paper so that $\frac{\hbar^{2}}{2\mu} = 1$.

\noindent

\section{\protect\bigskip Nodes, Extrema and Sumrules}
We develop a formalism in this section which would be suitable for applications to either the radial
Schroedinger equation for a specific partial wave in the domain
$[0,\infty]$ or the full line $[-\infty,+\infty]$. We consider the
solutions to
\begin{equation}
\frac{d^{2}}{dr^{2}}\Psi_{j}\ =\ \left( V\ -\ E_{j}\right) \ \Psi_{j}
\label{}
\end{equation}
satisfying the boundary conditions at the lower and upper end points of the domain at
$x_0$ and $x_1$
\begin{equation}
Lt_{r\to x_0}\Psi_{j}\ \to 0, \ \ Lt_{r\to x_1} \Psi_{j}\ \to \ 0
\label{}
\end{equation}
for the normalised eigenstates $\Psi_{j}$ corresponding to the eigenvalues $E_{j}$.
The state with $j=1$ corresponds to the groundstate with no nodes and the state
$\Psi_{j}$ has $(j-1)$ nodes. Let the nodes of the eigenstate $\Psi_{n}$
be at $r=R_0$ and let the outermost node be at $r={\tilde R}_{0}$. Then $\Psi_{n}$ has no nodes in the domain ${\tilde R}_{0}\ <\ r\ <\
x_1$. The second linearly independent solution at
$E_{n}$ is then given by
\begin{equation}
{\tilde \Psi}_{n}\left(r\right)\ =\ \Psi_{n}\left(r\right)\ \int_{R}^{r} \frac{1}{\Psi_{n}^{2}\left(y\right)}\ dy
\label{}
\end{equation}
where $R$ is a constant which may be chosen according to some
appropriate requirement. If we choose $R > {\tilde R}_0$ then in the domain $R\
 <\ r\ <x_1$ there can be no infinities arising from the denominator inside
the integral and ${\tilde \Psi}_{n}$ is well defined in this domain. The Wronskian
relation
\begin{equation}
\Psi_{n}\ \frac{d}{dr} {\tilde \Psi}_{n}\ -\ {\tilde \Psi}_{n}\
{\frac{d}{dr}}\Psi_{n}\ =\ 1 \label{}
\end{equation}
shows that $Lt_{r\to R_0}\ {\tilde \Psi}_{n}\ \ne \ 0$ but has a
finite value determined by the derivative of $\Psi_{n}$ at $r=R_0$. 

\noindent

\bigskip

The differential equation satisfied by the eigenstates may be used to
represent the Wronskian between the states $j$ and $k\ (j \ne k)$ in
terms of the overlap integrals between the different orthonormal eigenstates in the
form 
\begin{equation}
\int_{x_0}^{r} \Psi_k \left(y\right)\ \Psi_j \left(y\right)\ dy\ =\ \int_{x_1}^{r} \Psi_{k}\left(y\right)\ \Psi_{j}\left(y\right)\ dy\
=\ \frac{\left(\Psi_{k}\ {\dot \Psi}_{j}\ -\ {\dot \Psi}_{k}\ \Psi_{j}\right)}{\left( E_{k}\ -\ E_{j}\right)}  \label{}
\end{equation}
where the dot denotes a derivative with respect to $r$. 
If we now define
\begin{equation}
G\left(r,\tilde r\right)\ =\ \sum_{j\ne n} \frac{\Psi_{j}\left(r\right)\
\Psi_{j}\left(\tilde r\right)}{\left(E_{n}\ -\ E_{j}\right)} \label{}
\end{equation}
where the sum is over a complete set of eigenstates excluding the state
$n$, then using the equality in eq. (5) it can be established that
\begin{equation}
\left(\Psi_n (r) \frac{\partial}{\partial r}\ - {\dot \Psi}_n (r)\right)\ G\left(r,\tilde r\right)=\ \sum_{j\ne
n}\Psi_{j}\left(\tilde r\right)\ \int_{x_1}^{r} \Psi_{n}\left(y\right)\
\Psi_{j}\left(y\right)\ dy \ .\label{}
\end{equation}
The Green's functions $G(r,{\tilde r};E)$ considered in the usual
textbooks (Morse and Feshbach[1], for example) are constructed at
energies $E$ which are not one of the eigenenergies $E_n$. In contrast the
function $G$ considered here is constructed with $E=E_n$. Using the completeness relation satisfied by the eigenstates 
\begin{equation}
\sum_{j\ne n}\Psi_j (\tilde r)\ \Psi_j (y)\ =\ \delta (\tilde r - y)\
-\ \Psi_n (\tilde r)\ \Psi_n (y) \label{}
\end{equation}
it can be shown that
\begin{equation}
\left(\Psi_n (r) \frac{\partial}{\partial r} - {\dot \Psi}_n (r)\right) G\left(r,\tilde
r\right) = -\Psi_{n}\left(\tilde r\right)
 \left(\theta\left(r -
\tilde r\right)\int_{x_{1}}^{r} + \theta\left(\tilde r
- r\right)\int_{x_{0}}^{r} \right)\Psi_{n}^{2} dy \label{}
\end{equation}
where $\theta (z)$ is the unit step function which vanishes when $z<0$
and has value 1 when $z>0$. 

\noindent

\bigskip

\subsection{\noindent Sumrules at nodes of $\Psi_n$}
Various interesting relations follow from
the differential equation (9). If we choose $r = R_0$ where $R_0$ is
a node of the state $\Psi_n$ at which $\Psi_{n}(R_0)=0$ then we get the relation
\begin{equation}
\sum_{j \ne n}\frac{\Psi_{j}\left(R_{0}\right)
 \Psi_{j}\left(\tilde r\right)}{\left(E_{n} - E_{j}\right)} =
 \frac{\Psi_{n}\left(\tilde r\right)}{{\dot \Psi}_{n}\left(R_{0}\right)}
 \left(\theta \left(R_0 -\tilde r \right) \int_{x_1}^{R_{0}}  + \theta\left(\tilde r - R_0\right) \int_{x_0}^{R_0} \right)\Psi_{n}^{2} dy
 .\label{}
\end{equation}
In particular setting $\tilde r = R_{0}$ in eq. (10) leads to
\begin{equation}
\sum_{j \ne n}\frac{\Psi_{j}^{2}\left(R_{0}\right)}{\left(E_{n}\ -\
E_{j}\right)}\ =\ 0 \ . \label{}
\end{equation}
Squaring both sides of eq. (10), integrating over the variable
$\tilde r$ in the full range $[x_0 , x_1]$ and using the orthonormality of the
states $\Psi_j$ it may be shown that
\begin{equation}
\sum_{j\ne n} \frac{\Psi_{j}^{2}\left(R_0\right)}{\left(E_n - E_j
\right)^{2}}\ =\ \frac{1}{{\dot \Psi}_{n}^{2}\left(R_0\right)}\int_{x_0}^{R_0}\Psi_{n}^{2} dy
\int_{R_0}^{x_1} \Psi_{n}^{2} dz
\ . \label{}
\end{equation}
At any node of any eigenstate the rest of the eigenstates
must fulfill the conditions implied by eqs. (11) and (12).

\noindent

\bigskip

\subsection{\noindent Sumrules at extrema of $\Psi_n$}
Another special case of of eq. (9) arises when $r=R_{1}$ where $R_{1}$ is an extremum
of $\Psi_{n}$ which satisfies ${\dot \Psi}_{n}\left(R_{1}\right) = 0$.
Eq. (9)  simplifies to
\begin{equation}
\sum_{j \ne n}\frac{{\dot
\Psi}_{j}\left(R_{1}\right)\Psi_{j}\left(\tilde r\right)}{\left(E_{n}\
-\ E_{j}\right)} = \frac{\Psi_{n}\left(\tilde
r\right)}{\Psi_{n}\left(R_{1}\right)}\left(\theta\left(R_1 -
\tilde r\right)\int_{R_{1}}^{x_{1}}- 
\theta\left(\tilde
r - R_1\right)\int_{x_0}^{R_{1}}\right)\Psi_{n}^{2} dy
\label{}
\end{equation}
which when differentiated with respect to $\tilde r$ and evaluated at 
$\tilde r =R_{1}$ leads to the identity
\begin{equation}
\sum_{j \ne n} \frac{{\dot
\Psi}_{j}^{2}\left(R_{1}\right)}{\left(E_{n}-E_{j}\right)}\ =\ -\delta(\tilde
r
- R_1)|_{\tilde r=R_1} \  .
\label{}
\end{equation}
Using the completeness relation of the eigenstates the above
relation may also be given in the form
\begin{equation}
\sum_{j \ne n} \left( \frac{{\dot
\Psi}_{j}^{2}\left(R_1\right)}{\left(E_{n}-E_{j}\right)}\ +\
\Psi_{j}^{2}\left(R_{1}\right)\right)\ =\ -\
\Psi_{n}^{2}\left(R_{1}\right) \ . \label{}
\end{equation}
Squaring both sides of eq. (13), integarting over the variable $\tilde
r$ in the full range $[x_0 , x_1]$ and using the orthonormality of the
states $\Psi_j$ it can be shown that
\begin{equation}
\sum_{j\ne n}\frac{{\dot \Psi}_{j}^{2}\left(R_1\right)}{\left(E_n -
E_j \right)^{2}}\ =\ \frac{1}{\Psi_{n}^{2}\left(R_1\right)}\int_{x_0}^{R_1}\Psi_{n}^{2} dy
\int_{R_1}^{x_1}\Psi_{n}^{2}dz\ . \label{}
\end{equation}
At any extremum of any eigenstate the derivative of all other
eigenstates must satisfy the conditions implied by 
 eqs. (15) and (16). 

\noindent

\bigskip

\subsection{\noindent Integral realtions valid outside the last node of
$\Psi_n$}
The differential equation (9) satisfied by the function $G$ defined
by eq. (6) 
is a first order differential equation which can be brought to the form
\begin{equation}
\frac{\partial }{\partial
r}\frac{G\left(r,\tilde r\right)}{\Psi_{n}\left(r\right)}\ =
\frac{\Psi_{n}\left(\tilde r\right)}{\Psi_{n}^{2}\left(r\right)}
\left(\theta\left(r - \tilde r\right)\int_{r}^{x_1}\Psi_{n}^{2} dz - \theta\left(\tilde r - r\right)\int_{x_0}^{r} \Psi_{n}^{2} 
 dz \right) \label{}
\end{equation}
and can be integrated from an arbitrary point $r_{2}$ 
 to get
the relation
\begin{equation}
\frac{G\left(r,\tilde r\right)}{\Psi_{n}\left(r\right)} -
\frac{G\left(r_{2},\tilde r\right)}{\Psi_{n}\left(r_{2}\right)} =
\Psi_{n}\left(\tilde r\right)  \int_{r_{2}}^{r}\frac{dy}{\Psi_{n}^{2}}\
\left( \theta\left(y-\tilde
r\right) \int_{y}^{x_1} - \theta\left(\tilde r - y\right)\int_{x_0}^{y}\right)\Psi_{n}^{2} dz
.\label{}
\end{equation}

\noindent

\bigskip

It is also possible to establish a differential equation for
 $G\left(r,r\right)$. Using eqs. (5) and (6) and
the limiting value $Lt_{z \to 0}\ \theta\left(z\right) = 1/2$ it
can be established that
\begin{equation}
S\left(r\right)\ \equiv\ G\left(r,r\right)\ =\ \sum_{j \ne n}
\frac{\Psi_{j}^{2}\left(r\right)}{\left(E_{n}-E_{j}\right)} \label{}
\end{equation}
satisfies
\begin{equation}
\Psi_{n}^{2}\frac{d}{dr}\frac{S}{\Psi_{n}^{2}}\ =\
\left(\int_{r}^{x_1} \Psi_{n}^{2}\  dy \ -\ \int_{x_0}^{r} \Psi_{n}^{2}
\ dy \right) \ . \label{}
\end{equation}
This equation can be integrated from any point $r_{2}$ to give
\begin{equation}
S\left(r\right) = \frac{\Psi_{n}^{2}\left(r\right)}{\Psi_{n}^{2}\left(r_{2}\right)} S\left(r_{2}\right) + 
\Psi_{n}^{2}\left(r\right)\left(\int_{r_{2}}^{r}
\frac{dy}{\Psi_{n}^{2}}\left(\int_{y}^{x_1}\Psi_{n}^{2} dz -
\int_{x_0}^{y}\Psi_{n}^{2} dz\right) \right) \ . \label{}
\end{equation}

\noindent

\bigskip

Different choices of $\tilde r$ and $r_{2}$ in eq. (18) lead
to different integral relations. If we set ${\tilde r} = r_2$ in eq. (18)
then the resulting expression can be rearranged to give
\begin{equation}
\frac{G\left(r,r_{2}\right)}{\Psi_{n}\left(r\right)\Psi_{n}\left(r_{2}\right)} -
\frac{G\left(r_{2},r_{2}\right)}{\Psi_{n}^{2}\left(r_{2}\right)}
= \int_{r_{2}}^{r}\frac{dy}{\Psi_{n}^{2}(y)}\left(\theta\left(y -
r_{2}\right)\int_{y}^{x_1} - \theta\left(r_{2} - y\right)
\int_{x_0}^{y} \right) \Psi_{n}^{2}(z) dz .\label{}
\end{equation}
By interchanging the labels $r$ and $r_{2}$ another relation like the one
given above may be derived and by addition of the two
relations we can show that
\begin{equation}
\sum_{j \ne n} \frac{1}{E_{n}-E_{j}}
\left(\frac{\Psi_{j}\left(r\right)}{\Psi_{n}\left(r\right)} -
\frac{\Psi_{j}\left(r_{2}\right)}{\Psi_{n}\left(r_{2}\right)}\right)^{2}
= - \int_{r_{<}}^{r_{>}} \frac{dy}{\Psi_{n}^{2}(y)} \label{}
\end{equation}
where $r_{<}$ ($r_{>}$) is the smaller (larger) of ($r,r_{2}$). The
integrand in eq. (23) is free of singularities in the domain of
integration if both $r$ and $r_2$ are greater than the last node
${\tilde R}_0$ of $\Psi_{n}$.

\noindent

\bigskip
Various integral relations follow from eq. (23). For example
multiplying eq. (23) by $\Psi_{n}^{2}\left(r\right) 
\Psi_{n}^{2}\left(r_{2}\right)$, integrating over both the variables
from ${\tilde R}_0$ to $x_1$ and using the notation
\begin{equation}
A_{jk} = \int_{{\tilde R}_0}^{x_1} \Psi_{j}\left(y\right) \ \Psi_{k}\left(y\right)\ dy
\label{}
\end{equation}
and noting that when $r=R_0$ is a node of $\Psi_{n}$ eq. (5) gives 
\begin{equation}
A_{nj}\ =\ \frac{{\dot
\Psi}_{n}\left(R_0\right)\Psi_{j}\left(R_0\right)}{\left(E_{n}\ -\
E_{j}\right)} \ ,\ j\ne n\ ,\label{}
\end{equation}
we can establish that 
\begin{align}
A_{nn}\sum_{j \ne n} \frac{A_{jj}}{\left(E_{n}-E_{j}\right)} - {\dot
\Psi}_{n}^{2}\left({\tilde R}_0\right)\sum_{j\ne n}\frac{\Psi_{j}^{2}\left({\tilde R}_0\right)}{\left(E_{n}-E_{j}\right)^{3}} &= -\int_{{\tilde R}_0}^{x_1}
\Psi_{n}^{2}(r) dr \int_{r}^{x_1}\frac{dy}{\Psi_{n}^{2}(y)}
\int_{y}^{x_1}\Psi_{n}^{2}(z) dz  \notag \\
&= -\int_{{\tilde R}_0}^{x_1}\frac{dr}{
\Psi_{n}^{2}(r)}\int_{{\tilde R}_0}^{r}\Psi_{n}^{2}(y)dy\int_{r}^{x_1}\Psi_{n}^{2}(z) dz   . \label{}
\end{align}
A special case of the above relation arises if we consider the
groundstate with $n = 1$ for which ${\tilde R}_0 = x_0$ and for
all the eigenstates $\Psi_{j}\left({\tilde R}_0\right) = 0$. We thus get the
sum rule
\begin{equation}
\sum_{j=2}^{\infty}\frac{1}{\left(E_{1}-E_{j}\right)} =
-\int_{x_0}^{x_1}\Psi_{1}^{2}\left(r\right) dr\int_{r}^{x_1}\frac{dy}{\Psi_{1}^{2}\left(y\right)}\int_{y}^{x_1}\Psi_{1}^{2}\left(z\right) dz \label{}
\end{equation}
which expresses the inverses of the separation of the eigenvalues of a
confining potential from
the groundstate eigenvalue in terms of an integral over the nodeless
groundstate eigenfunction.

\noindent

\bigskip

The main results derived in this  paper are the relations expressed in
eqs. (9), (11), (12), (15), (16), (23), (26) and (27). 
 In the following sections exactly solvable examples will be used to
illustrate the sum rules derived in this section.

\noindent

\section{\protect\bigskip Examples of sumrules at nodes and extrema}

\subsection{\protect\bigskip Particle in a Box}

In this section we consider the example of a free particle confined
in a box with infinite walls at $x_0 = 0$ and $x_1 = \pi$. The normalised
eigenfunctions and eigenenergies are given by
\begin{equation}
\Psi_{j}\left(r\right)\ =\ \sqrt{\frac{2}{\pi}}\ \sin jr \ \ , \ \
E_{j}\ =\ j^{2}\ ,\ j=1,2...  \ .\label{}
\end{equation}
There is a node of the eigenfunction $\Psi_{n}$ at $R_0
=\pi\left(n-1\right)/n $. We first examine the sum
\begin{equation}
G\left(r,\tilde r\right) = \frac{2}{\pi}\sum_{j \ne n} \frac{\sin jr \sin j\tilde
r}{\left(n^{2}\ -\ j^{2}\right)} \label{}
\end{equation}
which can be simplified using partial fractions, addition formulae for
trigonometric functions and standard sums over sine functions (Gradshteyn
and Ryzhik 1965 [8]) to the form
\begin{equation}
G\left(r,\tilde r\right) = \frac{\sin nr \sin n\tilde r}{\pi n}
\left(-\frac{1}{2n} + r\cot nr + \tilde r \cot n\tilde r\ -\ \pi \cot
nr_{>} \right)  \label{}
\end{equation}
where $r_{>}$ is the larger of ($r,\tilde r$). Using 
\begin{equation}
\int_{0}^{R_{0}} \Psi_{n}^{2}\left(q\right) dq\ =\ \frac{R_{0}}{\pi} \ =\
\frac{n -1 }{n} \label{}
\end{equation}
and 
\begin{equation}
{\dot \Psi}_{n}\left(R_0\right)\ =\ \sqrt{\frac{2}{\pi}}\ n \cos \left(n
-1\right)\pi \ =\ (-1)^{n-1}\ n\ \sqrt{\frac{2}{\pi}} 
\label{}
\end{equation}
it is simple to show that for $r=R_0$ eq. (30) becomes
\begin{equation}
G\left(R_0,\tilde r\right) \ =
\ \frac{\Psi_{n}\left(\tilde r\right)}{{\dot \Psi}_{n}\left(R_0\right)}
\left(\theta\left(\tilde r - R_0\right)\frac{n-1}{n} - \theta\left(R_0 -
\tilde r\right) \frac{1}{n}\right)  \label{}
\end{equation} 
thereby verifying eq. (10).
For the choice ${\tilde r}=R_0 =\pi
- \pi/n$ eqs. (29) and (33) can be used to give
\begin{equation}
\sum_{j\ne n} \frac{\sin^{2} jR_0}{\left(n^2 -
j^2\right)}\ =\ 0 \label{}
\end{equation}
verifying eq. (11). Eq. (33) can be squared and integrated over ${\tilde
r}$ to show that 
\begin{equation}
\sum_{j\ne n} \frac{\sin^{2}jR_0}{\left(n^2 - j^2
\right)^2}\ =\ \frac{\pi^2}{4}\ \frac{n-1}{n^4} \label{}
\end{equation}
which is the sum rule arising from eq. (12) in this case.

\bigskip

We next examine 
\begin{equation}
{\dot G}\left(r,\tilde r\right) \ =\ 
 \frac{2}{\pi}\ \sum_{j \ne n} j
\ \frac{\cos jr\ \sin j\tilde r}{n^{2}\ -\ j^{2}} 
 \label{}
\end{equation}
which can be evaluated by taking the derivative of the relation in eq.
(30) with respect to $r$. There is an extremum of $\Psi_{n}$ at $R_{1}=
\pi - \pi/\left(2n\right)$. Hence
\begin{equation}
{\dot G}\left(R_{1},\tilde r\right) = \frac{\sin n\tilde r}{\pi}\left(-R_{1}\csc
nR_{1}\ +\ \pi \csc nR_{1}\ \theta\left(R_{1}-\tilde r\right)\right) . 
\label{}
\end{equation}
Using 
\begin{equation}
\int_{0}^{R_1} \Psi_{n}^{2} dy\ =\ \frac{R_1}{\pi} \label{}
\end{equation}
it can be shown that 
\begin{equation}
-\theta\left(\tilde r - R_{1}\right)\int_{0}^{R_{1}}\Psi_{n}^{2}dy +
\theta\left(R_{1} - \tilde r\right)\int_{R_{1}}^{\pi} \Psi_{n}^{2} dy \ =\ 
\left(-\frac{R_{1}}{\pi} + \theta\left(R_{1} - \tilde r\right)\right)  \label{}
\end{equation}
which together with eq. (37) verifies eq. (13) at the last extremum of $\Psi_{n}$ at $
R_{1}$. Differentiation of eq. (37) with respect to $\tilde r$, 
evaluation at the point $\tilde r = R_1$ and use of the completeness
relation leads to
\begin{equation}
\frac{2}{\pi}\sum_{j \ne
n}j^{2}\frac{\cos^{2}jR_{1}}{\left(n^{2}\ -\ j^{2}\right)}\ =\ -
\ \frac{2}{\pi}\sum_j \sin^{2}jR_1\ .  \label{}  
\end{equation}
It is possible to prove this directly by starting from the equality in
 eq. (30) for $r=\tilde
r$ and show that for $R_1 =\pi -
\pi/\left(2n\right)$
\begin{align}
\frac{\pi}{2}\ G\left(R_1 ,R_1\right)\ &=\ \sum_{j\ne n}
\frac{\sin^{2}jR_1}{n^2 - j^2}\ =\ -\frac{1}{4 n^2} \ , \notag\\
\frac{\pi}{4} \left(\frac{\partial^2}{\partial r^2} G\left(r,r\right)\right)|_{r=R_1}\
\ &=\ \sum_{j\ne n}\frac{j^2 \cos 2jR_1}{n^2 - j^2}\ =\  -\frac{3}{4} \label{}
\end{align}
which leads to the relation
\begin{equation}
\sum_{j \ne n} \left(\frac{j^2\cos^{2}jR_1}{\left(n^2 -j^2\right)}\ +
\ \sin^{2}jR_1\right)\ =\ -1\ =\ -\sin^{2}nR_1 \label{}
\end{equation}
thereby verifying eq. (15). By squaring eq. (37) and integrating over
${\tilde r}$ it can also be shown that
\begin{equation}
\sum_{j\ne n}\frac{j^2 \cos^2 jR_1}{\left(n^2 - j^2\right)^2}\ =\
\frac{\pi^2}{16}\ \frac{2n-1}{n^2} \label{}
\end{equation}
which is the sum rule arising from eq. (16) in the present case.

\bigskip

We next examine eq. (21) which in this example becomes
\begin{equation}
\Delta = \frac{G\left(r,r\right)}{\Psi_{n}^{2}\left(r\right)} -
\frac{G\left(r_{2},r_{2}\right)}{\Psi_{n}^{2}\left(r_{2}\right)} =
\int_{r_{2}}^{r} \frac{dy}{\sin^2 y}  \left(\int_{y}^{\pi} \sin^{2} nz\ dz
- \int_{0}^{y} \sin^{2} nz\ dz\right) .\label{}
\end{equation}
Using eq. (30) it can be shown that
\begin{equation}
\Delta \ =\ \frac{2r - \pi}{2n}\ \cot nr\ \ -\ \ \frac{2r_{2} -
\pi}{2n}\ \cot nr_{2} \label{}
\end{equation}
in agreement with the direct evaluation of the integral on the right
hand side of eq. (44). 

\bigskip

We next examine eq. (23) which in this example gives the relation
\begin{equation}
\frac{G\left(r,r\right)}{\Psi_{n}^{2}\left(r\right)} +
\frac{G\left(r_{2},r_{2}\right)}{\Psi_{n}^{2}\left(r_{2}\right)} -
2\frac{G\left(r,r_{2}\right)}{\Psi_{n}\left(r\right)\Psi_{n}\left(r_{2}\right)} = -\frac{\pi}{2} \int _{r_{<}}^{r_{>}}\frac{dy}{\sin^2 y} = \frac{\pi}{2n}\left(\cot nr_{>}  - \cot nr_{<}\right). \label{}
\end{equation}
Using eq. (30) to express the various terms on the left hand side of
eq. (46) it is easy to check that the sum of the terms on the left hand
side yields the expression on the right hand side of the equation. 

\bigskip

The triple integral on the right
hand side of eq. (27) for this example can be evaluated to give
\begin{equation}
\frac{2}{\pi}\int_{0}^{\pi} \sin^{2} x\ dx \ \int_{x}^{\pi}
\frac{dy}{\sin^2 y} \ \int_{y}^{\pi} \sin^{2} z\ dz \ =
\ \frac{3}{4} \label{}
\end{equation}
and using partial fractions it may be shown that
\begin{equation}
\sum_{j=2}^{\infty}\frac{1}{\left(1 - j^2\right)} \ =\ -\frac{3}{4}\label{}
\end{equation}
thus verifying eq. (27).

\noindent

\subsection{\protect\bigskip Simple Harmonic Oscillator}

We consider an oscillator potential $V=x^2$ in the range
$[-\infty,\infty]$ corresponding to a frequency
$\omega=2$.
The oscillator length parameter equals 1 in the
units we have used in this paper. The energy levels and the
eigenfunctions are given by
\begin{equation}
E_{j+1} = \left(2j+1\right) \ ,\ \Psi_{j+1} =\left(\frac{1}{\pi}\right)^{1/4}
\sqrt{\frac{1}{2^j j!}} \exp\left(-x^2/2\right)\ H_{j}\left(x\right),\ j=0,1,2,...\label{}
\end{equation}
where $H_j (x)$ are Hermite polynomials which satisfy
\begin{equation}
\frac{dH_{j}}{dx}\ =\ 2j\ H_{j-1}(x)\ ,\ H_{2j}(0)\ =\ (-)^j
\ \frac{\left(2j\right)!}{j!}\ ,\ H_{2j+1} (0)\ =\ 0 \ . \label{}
\end{equation}
We examine the sum rules arising from
the choice $n=2$ which corresponds to the first excited state with has a single node at $x=0$. All the antisymmetric states with even values of 
$j$ vanish at $x=0$ and the symmetric states corresponding to odd values
of $j$ have limiting values at $x=0$ given by
\begin{equation}
\Psi_{2j+1}^{2}\left(0\right)\ =\ {\sqrt\frac{1}{\pi}}
\ \left(\frac{1}{2^{2j}\left(2j\right)!}\right)
\ \left(\frac{\left(2j\right)!}{j!}\right)^2\ ,\ j=0,1,... \label{}
\end{equation}
where the first two factors on the right hand side arise from  the normalisation integrals of the
harmonic oscillator eigenfunctions (Pauling and Wilson 1935 [9]) and the last factor arises from the values of the even order
Hermite polynomials at $x=0$ (Abramowitz and Stegun 1965 [10]). Hence
\begin{align}
\sum_{j\ne 2}^{\infty} \frac{\Psi_{j}^{2}\left(0\right)}{E_2 - E_j}\ &=
\ \sqrt{\frac{1}{4\pi}}
\ \left(-\sum_{k=0}^{\infty} \frac{\left(2k\right)!}{\left(2^k
k!\right)^2}\ \frac{1}{2k-1}\right) \notag \\
&=\ \sqrt{\frac{1}{4\pi}}\  Lt_{z\to 1}\ \left(1 - z\right)^{1/2} \ =\ 0\label{}
\end{align} 
which verifies eq. (11) for $n=2$. 

\bigskip

We next examine
\begin{align}
\sum_{j\ne 2}^{\infty} \frac{\Psi_{j}^{2}\left(0\right)}{\left(E_2 -
E_j\right)^2}\ &=\ {\sqrt{\frac{1}{16\pi}}}\ \sum_{k=0}^{\infty}
\frac{\left(2k\right)!}{\left(2^k k!\right)^2}\ \frac{1}{\left(2k -
1\right)^2} \notag \\
&=\ {\sqrt\frac{1}{16\pi}}\ Lt_{z\to 1}\ \left(z\arcsin z + \left(1 -
z^2\right)^{1/2}\right)\  =\ \frac{\sqrt{\pi}}{8}. \label{}
\end{align}
The normalised eigenfunction for $n=2$ given by
\begin{equation}
\Psi_2\left(x\right)\ =\ \left(\frac{4}{\pi}\right)^{1/4}\ x\ \exp\left(-x^2
/2\right) \label{}
\end{equation}
can be used to show that
\begin{equation}
\frac{1}{{\dot \Psi}_{2}^{2}(0)}\ \int_{-\infty}^{0} \Psi_{2}^{2} dy
\ \int_{0}^{\infty} \Psi_{2}^{2} dz\ =\ \frac{\sqrt\pi}{8} \label{}
\end{equation}
which when considered together with eq. (53) verifies eq. (12) for the $n=2$ first excited state of the simple
harmonic oscillator. 

\bigskip

To examine the sum rule arising from extrema of eigenfunctions we
consider the groundstate $n=1$ which has an extremum at $x=0$. For all
the symmetric states corresponding to all odd values of $j$ the
derivative of the eigenfunction at $x=0$ vanishes and for the
antisymmetric states corresponding to even values of $j$ the derivative
at $x=0$ is given by 
\begin{equation}
{\dot\Psi}_{2j}^{2}\left(0\right) \ =\ \sqrt{\frac{1}{\pi}}
\ \frac{1}{2^{2j-1}}\ \frac{1}{\left(2j-1\right)!}
\ \left(\frac{\left(2j\right)!}{j!}\right)^2 \ ,\ j=1,2,...\ .\label{}
\end{equation}
Using the values of the eigenfunctions and their derivatives at $x=0$
given by eqs. (51) and (56) we can show that
\begin{align}
\sum_{j\ne 1}\left(\frac{{\dot\Psi}_{j}^{2}\left(0\right)}{E_1 - E_j}\ +
\ \Psi_{j}^{2}\left(0\right)\right)\ &=\ -\sqrt{\frac{1}{\pi}}\ \left(\sum_{k=0}^{\infty}\ -\  \sum_{k=1}^{\infty}\right)\ \frac{\left(2k\right)!}{\left(2^k k!\right)^2}
\notag\\
\ &= \ -\ \sqrt{\frac{1}{\pi}} \ =\ -\Psi_{1}^{2}\left(0\right) \label{}
\end{align}
thereby verifying eq. (15) for the groundstate of the oscillator.

\bigskip

We next consider
\begin{align}
\sum_{j=2}^{\infty}\frac{{\dot\Psi}_{j}^{2}\left(0\right)}{\left(E_1 - E_j\right)^2}
\ &= \ \sqrt{\frac{1}{4\pi}}\ \sum_{k=0}^{\infty}\frac{1}{\left(2^k k!\right)^2}
\ \frac{\left(2k\right)!}{\left(2k+1\right)} \notag \\
&= \ \sqrt{\frac{1}{4\pi}}\ Lt_{z\to 1}\ \left( \arcsin z\right)\ = 
\ \sqrt{\frac{\pi}{16}} \ .\label{}
\end{align}
It can be shown that for the normalised groundstate eigenfunction
\begin{equation}
\Psi_{1}\left(x\right) = \left(\frac{1}{\pi}\right)^{1/4} \exp\left(-x^2
/2\right)\ ,\ \frac{1}{\Psi_{1}^{2}\left(0\right)}\ \int_{-\infty}^{0} \Psi_{1}^{2} dy
\int_{0}^{\infty} \Psi_{1}^{2} dz\ =\ \frac{1}{4}\ \sqrt{\pi} \label{}
\end{equation}
which when taken together with eq. (58) verifies the sum
rule given by eq. (16) for the $n=1$ groundstate of the oscillator.

\bigskip

\section{\protect\bigskip Discussion}

In this paper sum rules which must be satisfied at
the nodes and extrema of boundstate eigenfunctions of confining
potentials have been established. The sum rules in eqs. (11), (12), (15)
and (16) have been verified for the case of a particle confined in a box
and also explicitly for the case of a simple harmonic oscillator in the ground or
first excited states. However the sum rules are valid for all states of
the oscillator and for all confining potentials. When scattering states
are present the expressions have to be modified by the addition of an
integral to take account of the contribution from the scattering states
to the sum over the contribution from the discrete states. 

\bigskip

We have shown that in the domain after the last node of
an eigenfunction the feature that the inverse of the eigenfunction is
singularity free may be used to establish a variety of relations between
the values of all the other eigenfunctions in this region and integrals
involving the nodeless eigenfunction.
We have illustrated the sum rules in eqs. (23) and (27) for the case of
a particle in a box for which the sums and integrals converge and can be
carried out analytically. For the harmonic oscillator the sum and
integral in eq. (27) do not converge. 

\bigskip

The antisymmetric wavefunction for two non-interacting identical fermions
moving in the same 
single particle potential $V$ such that one of them is in the state $\Psi_n$ and the other in
$\Psi_j$ is given by
\begin{equation}
\Phi_{nj}\left(r_1,r_2\right)\ =\ \sqrt{\frac{1}{2}}\ \left(\Psi_{n}\left(r_1\right)
\ \Psi_{j}\left(r_2\right)\ -\ \Psi_{n}\left(r_2\right)\
\Psi_{j}\left(r_1\right)\right) \ . \label{}
\end{equation}
The relationship in eq. (23) may also be given in terms of $\Phi_{nj}$ in the form
\begin{equation}
\sum_{j\ne n} \frac{\Phi_{nj}^{2}\left(r_1,r_2\right)}{E_n - E_j}\ =\
-\frac{1}{2}\ \Psi_{n}^{2}\left(r_1\right)\
\Psi_{n}^{2}\left(r_2\right)\ \int_{r_<}^{r_>}
\frac{dy}{\Psi_{n}^{2}}\label{}
\end{equation}
which sheds an interesting light on the sum rule in terms of joint
probabilty density of two-particle states.

\bigskip

Supersymmetric Quantum Mechanics may be used to interpret the integral
on the right hand side of eq. (26). If we consider a potential $\tilde
V$ which is identical to $V$ in the region outside the last node of $\Psi_n$ at ${\tilde R}_0$ but has
an infinite wall at the last node, then the boundstate solutions in $\tilde V$
must vanish at $r={\tilde R}_0$ and  as $r\to x_1$. The groundstate energy of $\tilde V$ 
must be ${\tilde E}_1 =E_n$ because $\Psi_n$ goes to zero at $r={\tilde R}_0$ and as $r\to x_1$ but has no nodes inbetween. $\Psi_n$ is the groundstate eigenfunction of $\tilde V$
but has to be renormalised to 1 in the region [${\tilde R}_0,x_1$]. Let
the other boundstate eigenvalues of $\tilde V$ satisfying boundstate
boundary conditions at ${\tilde R}_0$ and $x_1$ be ${\tilde E}_j , j=2,3,..$ .
A supersymmetric partner to the
potential $\tilde V$ constructed by the elimination of its groundstate
at ${\tilde E}_1=E_n$ is
\begin{equation}
{\tilde V}_1 \ =\ {\tilde V}\ -\ \frac{d^2}{dr^2}\ \ln \Psi_{n}\left(r\right)
,\ r> {\tilde R}_0 \ , \label{}
\end{equation}
which is free of singularities for $r> {\tilde R}_0$. This
construction which is based on the methods of Supersymmetric Quantum
Mechanics (Sukumar 1985 [11]) guarantees that the boundstate spectrum of ${\tilde
V}_1$ is identical to that of $\tilde V$ except for missing the
groundstate of $\tilde V$ at ${\tilde E}_1$ ({\it i.e}) ${\tilde V}_1$
has spectrum ${\tilde E}_j , j=2,3,...$.  
\ It may be shown that a solution at the energy $E_n$ in ${\tilde V}_1$ is ${\tilde
\Phi} = 1/\Psi_n$. From this solution two other solutions which
satisfy boundary conditions at ${\tilde R}_0$ and $x_1$ can be
constructed in the form
\begin{align}
{\tilde \Phi}_1\ &=\ \frac{1}{\Psi_n \left(r\right)}\ \int_{{\tilde
R}_0}^{r} \Psi_{n}^{2}\left(y\right)\ dy\ ,\ \ Lt_{r\to {\tilde R}_0}\ {\tilde
\Phi}_1\left(r\right)\ \to 0 , \notag\\
{\tilde \Phi}_2\ &=\ \frac{1}{\Psi_n \left(r\right)}\ \int_{x_1}^{r}
\Psi_{n}^{2}\left(y\right)\  dy\ ,\ \ Lt_{r\to x_1}\ {\tilde \Phi}_2 \left(r\right)\ \to 0
\ \label{}
\end{align}
with the Wronskian 
\begin{equation}
W\ =\ {\tilde \Phi}_1\ \frac{d}{dr}\ {\tilde \Phi}_2\ -\ {\tilde
\Phi}_2\ \frac{d}{dr}\ {\tilde \Phi}_1 \ =\ \int_{{\tilde R}_0}^{x_1}
{\Psi}_n^2(y)\ dy\ .\label{}
\end{equation}
These solutions may be used to construct a Green's function for the potential ${\tilde
V}$ given by
\begin{equation}
{\tilde G}_1\left(r,{\tilde r}>r\right) \ =\ \frac{{\tilde \Phi}_1 \left(r\right)\ {\tilde
\Phi}_2 \left({\tilde r}\right)}{W}\ ,\ \ Lt_{r\to {\tilde R}_0} {\tilde G}\ \to
0,\ \ Lt_{{\tilde r}\to x_1} {\tilde G} \to 0 \ \label{}
\end{equation}
and ${\tilde G}_1(r,\tilde r) = {\tilde G}_1(\tilde r
,r)$. The trace of this Green's function (Sukumar 1990 [3]) is related to the spectrum of
${\tilde V}_1$ by
\begin{equation}
\int_{{\tilde R}_0}^{x_1} {\tilde G}_1 \left(r,r\right) dr\ =\ \frac{1}{W}
\ \int_{{\tilde
R}_0}^{x_1}\frac{dr}{{\Psi}_{n}^{2}}\ \int_{{\tilde
R}_0}^{r}{\Psi}_{n}^{2}dy\ \int_{x_1}^{r}{\Psi}_{n}^{2}dz \ =
\ \sum_{j\ne 1}
\frac{1}{E_n - {\tilde E}_j}\ .\label{}
\end{equation}
Using eqs. (60), (61), (64) and (66)  
 it may be shown that 
\begin{align}
\sum_{j\ne n}\frac{1}{E_n - E_j} \int_{{\tilde R}_0}^{x_1} 
 \int_{{\tilde R}_0}^{x_1}  \Phi_{nj}^{2}\left(r_1,r_2\right)
 dr_1\ dr_2 \ & = W\ \int_{{\tilde R}_0}^{x_1} {\tilde G}_1\left(r,r\right)
\ dr\notag\\ 
\ &= \left(\ \int_{{\tilde R}_0}^{x_1} {\Psi}_n^2 (y)dy\right)\ \sum_{j\ne 1}\frac{1}{E_n - {\tilde E}_j}\label{}
\end{align}
expressing the trace of the Green's function for the Supersymmetric partner
potential ${\tilde V}_1$ with the energy spectrum $({\tilde E}_j,
j=2,3,..)$ in terms of two-particle densities in the
potential $V$ with the energy spectrum $(E_j, j=1,2,..n,...)$.

We conclude by reiterating that the sum rules expressed in eqs. (11),
(12), (15) and (16) must be satisfied at all the nodes and extrema of
the boundstate eigenfunctions of confining potentials in
1-dimension. Also the sum rule in eq. (23) for confining potentials is a key 
result derived in this paper. As noted before it is possible to extend the
sum rule to potentials which have scattering states by the addition of
an additional integral to include the contribution from the scattering
states in addition to the contribution from the discrete states which
are included in eq. (23). We have focussed attention on confining
potentials because of the existence of exactly solvable problems for
which the sum rules can be explicitly verified. We have verified the sum rules 
for two exactly solvable confining potentials. We have interpreted one of
the sum rules using the notion of
two particle densities and established a connection with the trace of
the Green's function of a Super Symmetric partner in Super Symmetric
Quantum Mechanics.

\bigskip

\section{\protect\bigskip References}

1. Morse P. and Feshbach H. 1953 {\it Methods of Theoretical Physics}
   (New York: McGraw-Hill) Vol 1, 791-811.

\noindent
2. Berry M.V. 1986 {\it J.Phys. A: Math. Gen.} {\bf 19} 2281.

\noindent
3. Sukumar C.V. 1990 {\it Am. J. Phys.} {\bf 58} 561.

\noindent
4. Voros A. 2000 {\it J.Phys A: Math. Gen.} {\bf 33} 7423.

\noindent
5. Bender C.M. and Boettcher S. 1998 {\it Phys. Rev. Let.} {\bf 80}
   5243.

\noindent
6. Mezincescu G.A. 2000 {\it J.Phys. A: Math. Gen.} {\bf 33} 4911.

\noindent
7. Bender C.M. and Wang Q. 2001 {\it J.Phys. A: Math. Gen.} {\bf 34} 3325.

\noindent
8. Gradshteyn I.S. and Ryzhik I.M. 1965 {\it Tables of Integrals,
Series and Products} (New York: Academic) 38.

\noindent
9. Pauling L. and Wilson E.B. 1935 {\it Introduction to Quantum
Mechanics} ( New York: McGraw-Hill) 80.

\noindent
10.Abramowitz M and Stegun I.A. 1965 {\it Handbook of Mathematical
Functions} (New York: Dover) 777.

\noindent
11.Sukumar C.V. 1985 {\it J.Phys A: Math. Gen.} {\bf 18} 2917.

\end{document}